\documentclass[referee]{raa}            

\usepackage{graphicx,times}

\begin{document}

\title{State transitions triggered by inverse magnetic field: probably applied in high-mass X-ray binaries?}

\volnopage{Vol.0 (200x) No.0, 000--000}      
\setcounter{page}{1}          

   \author{Shuang-Liang Li
      \inst{1}
   \and Zhen Yan
      \inst{1}
      }

      \institute{Key Laboratory for Research in Galaxies and
Cosmology, Shanghai Astronomical Observatory, Chinese Academy of
Sciences, 80 Nandan Road, Shanghai, 200030, China; {\it lisl@shao.ac.cn, zyan@shao.ac.cn}\\
   }

    \date{Received~~2009 month day; accepted~~2009~~month day}

\abstract{ Previous works suggested that the state transitions in an X-ray binary can be triggered by accreting inverse magnetic field from its companion star. A key point of this mechanism is the accretion and magnification of large-scale magnetic fields from outer boundary of a thin disk. However, how such a process can be realized is still an open question. In this work, we check this issue in a realistic X-ray binary system. According to our calculations, a quite strong initial magnetic field $B\sim 10^2-10^3$ G is required in order to assure that the large-scale magnetic field can be effectively dragged inward and magnified with the accretion of gas. Thus, such a picture probably can be present in high-mass X-ray binaries possessing strong stellar magnetic field, e.g., Cyg X-1.
\keywords{accretion, accretion discs -- black hole physics --
(magnetohydrodynamics) MHD -- X-rays: binaries}
}

\authorrunning{S.-L. Li \& Z. Yan}
\titlerunning{State transition in X-ray binaries}

\maketitle

\section{Introduction}

It is well known that black hole (BH) X-ray binaries (XRBs) have shown different X-ray spectral states (see reviews by Remillard \& McClintock~\cite{r2006}; Belloni~\cite{Belloni2010}; Zhang~\cite{z2013}). The two fundamental X-ray spectral states are the low/hard (LH) and high/soft (HS) states, where the X-ray emission in LH state dominated by a non-thermal component coming from a hot accretion flow (Narayan \& Yi~\cite{n1994}; Yuan \& Narayan~\cite{y2014}), and the X-ray emission in HS state is dominated by a thermal component coming from an optically thick and geometrically thin disk (Shakura \& Sunyaev~\cite{s1973}). During the transition between LH and HS states, both thermal and non-thermal components are important, which is usually called intermediate state. Though the mechanism for the transition between these two X-ray spectral states has been widely studied, it is still an open question so far.

Mass accretion rate has long been believed as the dominant parameter in determining spectral state transitions (Esin et al.~\cite{Esin1997}). In this kind of truncated disk model, the accretion geometry in the LH state is an inner hot accretion flow plus an outer thin disk (Meyer et al.~\cite{m2000}; Done et al.~\cite{Done2007}; Liu et al.~\cite{l2011}; Yuan \& Narayan~\cite{y2014}), where the hot accretion flow only survive below 0.01 $L_\mathrm{Edd}$ (Esin et al.~\cite{Esin1997}), or up to 0.1$L_\mathrm{Edd}$ by adopting an alternative solution in Yuan et al. (\cite{Yuan2007}). So the state transition occurs at a nearly constant luminosity (several percent of Eddington luminosity), which is inconsistent with the observation that the hard-to-soft transition luminosity varies up to two orders of  magnitude (Homan et al.~\cite{Homan2001}; Gierli$\acute{\rm n}$ski \& Done~\cite{Gierlinski2003}; Zdziarski et al.~\cite{Zdziarski2004}; Yu \& Yan~\cite{yu2009}). Therefore, many other parameters has been proposed. Such as, the Comptonizing region constrained from timing variability has been considered as the other independent parameter to determine the state transition behaviour (Homan et al.~\cite{Homan2001}); the recent accretion history may play an important role in determining the transition luminosity (Homan \& Belloni~\cite{Homan2005b}); the disk mass has been suggested as the initial condition in determining the state transitions (Yu et al.~\cite{Yu2004}; Yu \& Dolence~\cite{Yu2007}; Yu \& Yan~\cite{yu2009}); and the process of disc tearing could produce variety of behaviour capable of explaining state transitions (Nixon \& Salvesen~\cite{Nixon2014}). But there is no conclusive evidence for either argument, this subject still remains a mystery. In this work, we are going to discuss another promising parameter: magnetic field.

Large-scale magnetic fields, which are believed to accelerate the jet (see reviews of jet formation models by Yuan \& Narayan~\cite{y2014}), should be present in LH state. Recently, strong disk winds are found to exist in the HS state of some X-ray binaries (Ponti et al.~\cite{p2012}), which may also be driven by the ordered magnetic field. Thus, it is possible that  magnetic field can be another parameter affecting the state transitions. Igumenshchev (\cite{i2009}) and Dexter
et al. (\cite{d2014}) suggested that a magnetosphere could form in the inner region of disk by accumulating enough magnetic flux (Narayan et al.~\cite{n2003}; McKinney et al.~\cite{m2012}) and that the state transition could be triggered by accreting inverted magnetic field from companion star. The LH state can correspond to a truncated accretion disk with large truncation radius $R_{\rm tr}$. Recently, the advection of magnetic field from outer disk region is believed to be a promising way to form a large-scale field in accretion disk. Nevertheless, how it forms in a thin disk has been a long-standing debatable issue. Thus, the key point of this picture is the accretion and magnification of large-scale magnetic fields from outer boundary of a thin disk. Due to the low radial velocity and long advection timescale, the accretion of a standard thin disk had been found to be ineffective on magnifying the field (van Ballegooijen~\cite{v1989}; Lubow et al.~\cite{l1994}). However, Spruit \& Uzdensky (\cite{s2005}) suggested that the advection of field can be accelerated by some strong bundles of flux threading the disk because they can simultaneously increase the radial velocity by taking away the angular momentum of the disk and decrease the diffusion by suppressing the turbulence. Cao \& Spruit (\cite{c2013}) considered a thin disk with moderately weak large-scale magnetic field, where the angular momentum is totally transferred by disk winds, and found that the advection efficiency of field can be greatly increased too. This mechanism was revisited by considering the effects of winds on the disk structures in Li \& Begelman (\cite{l2014}). They found that the advection timescale of field can do be smaller than the diffusion timescale for the main reason that the disk temperature is greatly decreased because the winds take away most of the viscous dissipated energy, resulting in the decrease of the magnetic diffusivity $\eta$ and the increase of the diffusion timescale (see section \ref{models} for details). Except for the models mentioned above, the other mechanism to solve the field diffusion problem on the formation of large scale fields in a thin disk is the coronal mechanism (e.g., Rothstein \& Lovelace~\cite{r2008}, Beckwith et al.~\cite{b2009}), which suggested the field can be advected inwards through the corona region and thus avoided the diffusion problem. In this work, we try to check whether the large-scale magnetic field on a thin disk can be magnified in realistic X-ray binary systems.

\section{MODEL}\label{models}

We consider a realistic thin disk which accretes gas from the companion star in X-ray binaries. Whether the magnetic field can be effectively dragged inwards depends on the competition of advection timescale and diffusion timescale. If the advection timescale $\tau_{\rm adv}$ is far smaller than diffusion timescale $\tau_{\rm dif}$, the field can be magnified. The diffusion timescale $\tau_{\rm dif}$ can be given by $\sim RH \kappa_{0}/\eta$ (Cao \& Spruit~\cite{c2013}; Li \& Begelman~\cite{l2014}), where $\eta \sim \nu$ ($\nu$ is the viscosity coefficient) as suggested by recent MHD simulations (Fromang \& Stone~\cite{f2009}; Guan \& Gammie~\cite{g2009}). When the magnetic torque is far larger than the viscous torque and thus dominates the transportation of disk angular momentum, the strong disk winds driven by magnetic torque will take away lots of energy released in the disk, which results on the decrease of $\nu$ ($\sim \alpha c_{\rm s} H$) and the increase of diffusion timescale. Therefore, the field can be effectively magnified when the magnetic torque dominates over the viscous torque.

Consider a thin disk with very weak field, the magnetic and viscous torques are given by $T_{\rm m}=B_{\rm p}B_{\rm \phi}R/2\pi$ (Livio et al.~\cite{l1999}; Cao~\cite{c2002}) and $W_{\rm R\phi}=2H\alpha P_{\rm tot}$, respectively. Thus, the ratio of magnetic torque and viscous torque is
\begin{equation}
\frac{T_{\rm m}}{W_{\rm R\phi}}\sim \frac{0.2R}{H\alpha \beta_{\rm p}}\sim\frac{130}{\beta_{\rm P}}
\end{equation}
with $B_{\rm \phi}=0.1 B_{\rm p}$ and $\alpha=0.1$, where $\beta_{\rm p}=(P_{\rm {gas}}+P_{\rm rad})/(B_{\rm p}^2/8\pi)$ and $H/R\sim 1.6 \times 10^{-2}$ are adopted. Where the disk scale height $H$ is given by
\begin{eqnarray}
H&=&1.5\times 10^{3} \alpha^{-1/10} \dot{m}^{3/20}m^{9/10}r^{9/8}(1-r^{-1/2})^{3/20}\nonumber\\
&=&4.7\times 10^{8} {\rm cm},
\end{eqnarray}
where $r=R/R_{\rm g}$, $R=10^4 R_{\rm g}$, $\dot{m}=1$ ($\dot{m}=\dot{M}/\dot{M}_{\rm crit}, \dot{M}_{\rm crit}=1.5\times 10^{17}m {\rm gs^{-1}}$, $m=M/M_{\odot}$) and $m=10$ (Kato et al.~\cite{k1998}). Therefore, if the field is strong enough, e.g., $\beta_{\rm p}\sim 1-10$, the magnetic torque will dominate the transportation of disk angular momentum. For a standard thin disk with the same disk parameters, the disk pressure is dominated by gas pressure and can be given approximately by
\begin{eqnarray}
P_{\rm gas}&\simeq& 3.12\times 10^{17} \alpha^{-9/10} \dot{m}^{17/20}m^{-9/10}R^{-21/8}\nonumber\\
&&\times(1-R^{-1/2})^{17/20}\nonumber\\
&=&1.38\times10^6 {\rm g cm^{-1} s^{-2}}
\end{eqnarray}
(Kato et al.~\cite{k1998}). However, as pointed out by Li~\cite{l2014b}, the pressure of a disk could be $10^{2}$ times smaller than that of a standard disk for the reason that the most energy released in the disk is taken away by disk winds, resulting in the decrease of gas pressure (see Figure 2 therein). Thus, the disk pressure at outer radius will reduce to $\sim 1.38\times10^{4} {\rm g cm^{-1} s^{-2}}$ for a thin disk with strong disk winds. For $B=10^3 $ G, the corresponding magnetic pressure is $P_{\rm m}=B_{\rm p}^2/8\pi\simeq4\times 10^4 {\rm g cm^{-1} s^{-2}}$, which corresponds to $\beta_{\rm p} < 1$ when $B_{\rm \phi}=0.1 B_{\rm p}$ is adopted.

Thus, a field strength $B \sim 10^2-10^3$ G from the outer boundary of a thin disk is required in order to assure the dominated role of magnetic torque. It is reasonable to assume that the accreting gas at outer boundary carries similar magnetic field with the companion star. Therefore, the scenario of this model looks probable in the systems with high magnetic field companion stars. So far as we know, most the Galactic BH XRB systems harbor a low mass companion star. Among those BH low-mass X-ray binaries (LMXBs) with solid measurements, most of the companion stars are K-type stars (Ritter \& Kolb~\cite{Ritter2003}; Casares \& Jonker~\cite{Casares2014}). We found the magnetic field strength of different spectral type stars in the catalog of Bychkov et al. (\cite{Bychkov2009}) are roughly consistent with a normal distribution in logarithmic scale. The average magnetic field strength of K-type stars  is $\sim 20$ G, which is smaller than the requirement of Igumenshchev (\cite{i2009})'s model according to our calculations. Cygnus X-1 is the only ever known Galactic BH XRB harbor a high mass companion star, the spectral type of which is O9.7 (Bolton~\cite{Bolton1972}; Orosz et al.~\cite{Orosz2011}). The average magnetic field strength of the O-type stars in Bychkov et al. (\cite{Bychkov2009}) is $\approx 340$ G. Fortunately, Karitskaya et al. (\cite{Karitskaya2010}) had measured the $<B_{z}> \sim 600$ G at the outer radius of the accretion disk (~$2\times10^{5} R_\mathrm{g}$ ), which is higher enough to satisfy the condition $\beta_{\rm p}< 10$. Therefore, we think this model is promising in explaining the spectral state transition in high-mass X-ray binaries possessing strong stellar magnetic field, such as, Cyg X-1.

\section{SUMMARY AND DISCUSSION}\label{DISCUSSION}

In this work, we investigate whether the ordered magnetic field can be magnified from the outer boundary of a thin disk, which is the key point to realize the mechanism that state transition can be triggered by accreting inverse magnetic field from companion star. According to our calculations, a quite strong initial magnetic field $B\sim 10^2-10^3$ G is required in order to assure the dominant role of magnetic torque on transferring the angular momentum. Thus, such a picture probably can be present in some high-mass X-ray binaries.

\begin{figure}
\centering
\includegraphics[width=8cm]{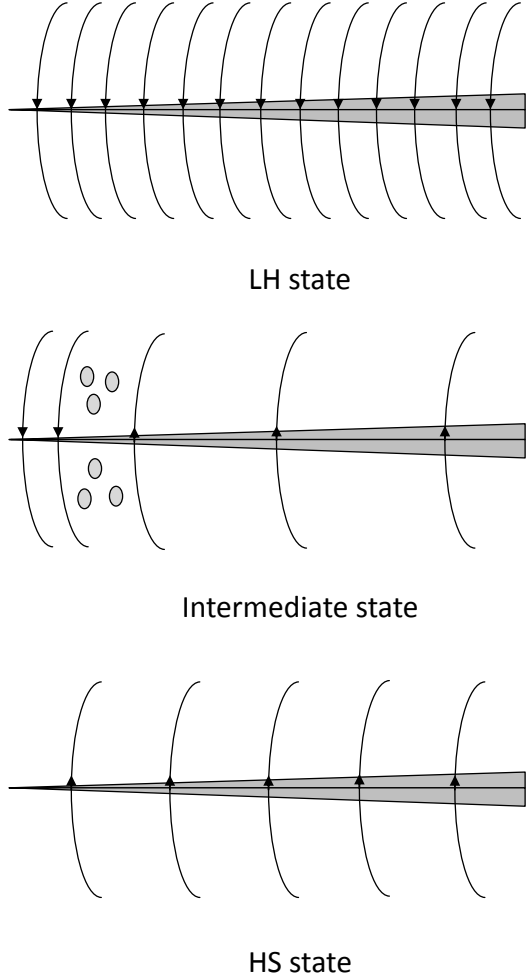}
\caption{Schematic diagram for the magnetic field threading on the disk in state transitions. \label{f1}}
\end{figure}

Interestingly, as presented in Li (\cite{l2014b}), the disk winds driven by large-scale magnetic field can take away most of the energy released in the disk and thus help to cool the disk. The radiative efficiency of a thin disk with winds could be $10^{3}$ times smaller than that of a standard thin disk. Thus, the thermal component of a disk will disappear if there are strong disk winds (see Figure \ref{f1}). Such a state seems can correspond to the LH state in X-ray binaries, which is similar to that suggested in Livio et al. (\cite{l2003}). The system will gradually turn into the intermediate state once the accretion of opposed field starts. The transient jets in intermediate state can come from the acceleration from reforming field to the bubbles/outflows, which is produced by the magnetic reconnection (Igumenshchev~\cite{i2009}; Dexter et al.~\cite{d2014}; Khiali et al.~\cite{k2015}), or from the magnetic rope where the energy reaches a threshold (Yuan et al.~\cite{y2009}). With the accretion of inverted field, the former large-scale field will gradually vanish due to the magnetic annihilation with the inverted field. After the disappearance of original field and before the reformation of inverted large-scale field, the thin disk will become radiatively efficiently again and enter HS state. Wind can be easily launched from the surface of a radiatively inefficient accretion flow due to the positive Bernoulli parameter of the gas (Narayan \& Yi~\cite{n1994}; Blandford \& Begelman~\cite{b1999}; Yuan et al.~\cite{y2012}; Gu~\cite{g2015}). However, other conditions, e.g., radiation pressure and/or large-scale magnetic field are required in order to drive winds from a thin disk (Murray et al.~\cite{m1995}; Blandford \& Payne~\cite{b1982}). In the case of ordered magnetic field, the inclination angle of field lines with respect to the surface of the disk is required to be smaller than $60^{\circ}$ in order to launch winds from a cold thin disk (Blandford \& Payne~\cite{b1982}). But radiation force can help to realize this process as suggested by Cao (\cite{c2012}, \cite{c2014}). Strong disk winds do be validated by the discovery of highly ionized absorbers in the HS states of some X-ray binaries (Ponti
et al.~\cite{p2012}), which may come from the driving of large-scale magnetic field.

The typical magnetic flux of a thin disk threading by large-scale field is $\Phi \sim 10^{22} $ G cm$^{-2}$ as suggested by Igumenshchev (\cite{i2009}). With the presence of large-scale field, the radial velocity of a thin disk is about $V_{\rm R} \sim10^5$ ${\rm cm s^{-1}}$ at $R\sim 10^4 R_{\rm g}$, which is one-two orders of magnitude larger than that of a standard thin disk (Li \& Begelman~\cite{l2014}). Thus, the timescale to accumulate the observed magnetic flux is $\tau \sim \Phi/B V_{\rm R}R \sim 10^{4-6}$ s for a field strength $B\sim 10^{2-3}$ G. The observed transition timescales between LH and HS states in Cyg X-1 are several days (Grinberg et al.~\cite{g2013}), which is roughly consistent with the results of our model. In order to avoid the formation of a magnetically-arrested accretion flow (MAD, McKinney et al.~\cite{m2012}), a balance of magnetic field advection and diffusion should be required. However, how such a balance can be achieved and kept stable is still unclear (e.g., Bisnovatyi-Kogan \& Lovelace~\cite{b2012}; Cao \& Spruit~\cite{c2013}). If the answer is not, an MAD will be built naturally in the inner disk region. In such a case, the flux accumulated timescale $\sim 10^4-10^6$ s is also roughly consistent with the timescale of an outburst in X-ray binaries.

\section* {ACKNOWLEDGEMENTS}
We thank the referee for his/her very helpful report. Shuangliang Li thanks F. Yuan and X. Cao for helpful comments and discussion. This work is supported by the NSFC (grants 11233006, 11373056) and the Science and Technology Commission of Shanghai Municipality (13ZR1447000). Zhen Yan acknowledges the support from the Knowledge Innovation Program of the Chinese Academy of Sciences and National Natural Science Foundation of China under grant No. 11403074.

\clearpage


\begin{thebibliography}{}





\bibitem[2009] {b2009} Beckwith, K., Hawley, J. F., \& Krolik, J. H. 2009, ApJ, 707, 428

\bibitem[2010] {Belloni2010} Belloni, T. M. 2010, in Lecture Notes in Physics, Berlin Springer Verlag, ed. T. Belloni, Vol. 794, 53

\bibitem[2012] {b2012} Bisnovatyi-Kogan, G. S., \& Lovelace, R. V. E. 2012, ApJ, 750, 109


\bibitem[1999]{b1999} Blandford R. D., Begelman M. C., 1999, MNRAS, 303, L1

\bibitem[1982]{b1982} Blandford, R. D., \& Payne, D. G. 1982, MNRAS, 199, 883

\bibitem[1972] {Bolton1972} Bolton, C. T. 1972, NatPS, 240, 124

\bibitem[2009] {Bychkov2009} Bychkov, V. D., Bychkova, L. V., \& Madej, J. 2009, MNRAS, 394, 1338

\bibitem[2002]{c2002} Cao, X., 2002, MNRAS, 332, 999

\bibitem[2012]{c2012} Cao, X., 2012, MNRAS, 426, 2813

\bibitem[2014]{c2014} Cao, X., 2014, ApJ, 783, 51

\bibitem[2013]{c2013} Cao, X., \& Spruit, H. C., 2013, ApJ, 765, 149

\bibitem[2014]{Casares2014} Casares, J., \& Jonker, P. G. 2014, SSR, 183, 223

\bibitem[2014]{d2014} Dexter, J., McKinney, J. C., Markoff, S., \& Tchekhovskoy, A. 2014, MNRAS, 440, 2185

\bibitem[2007] {Done2007} Done, C., Gierli{\'n}ski, M., \& Kubota, A. 2007, A\&AR, 15, 1



\bibitem[1997] {Esin1997} Esin, A. A., McClintock, J. E., \& Narayan, R. 1997, ApJ, 489, 865

\bibitem[2009] {f2009} Fromang, S., \& Stone, J. M. 2009, A\&A, 507, 19


\bibitem[2003]{Gierlinski2003} Gierli{\'n}ski, M., \& Done, C. 2003, MNRAS, 342, 1083

\bibitem[2013]{g2013} Grinberg V. et al., 2013, A\&A, 554, A88

\bibitem[2015]{g2015} Gu, Wei-Min, 2015, ApJ, 799, 71

\bibitem[2009]{g2009} Guan, X., \& Gammie, C. F. 2009, ApJ, 697, 1901

\bibitem[2005]{Homan2005b} Homan, J., \& Belloni, T. 2005, Ap\&SS, 300, 107

\bibitem[2001] {Homan2001} Homan, J., Wijnands, R., van der Klis, M., et al. 2001, ApJS, 132, 377

\bibitem[2009]{i2009} Igumenshchev, I. V. 2009, ApJL, 702, L72


\bibitem[2010]{Karitskaya2010} Karitskaya, E. A., Bochkarev, N. G., Hubrig, S., et al. 2010, IBVS, 5950, 1

\bibitem[1998]{k1998} Kato S., Fukue J., Mineshige S., eds, 1998, Black-Hole Accretion Disks, Kyoto Univ. Press, Kyoto, Japan


\bibitem[2015]{k2015} Khiali, B., de Gouveia Dal Pino, E. M. \& del Valle, M. V. 2015, MNRAS, 449, 34


\bibitem[2014]{l2014b} Li S.-L., 2014, ApJ, 788, 71

\bibitem[2014]{l2014} Li S.-L., \& Begelman, M. C., 2014, ApJ, 786, 6

\bibitem[2011]{l2011} Liu, B. F., Done, C., \& Taam, R. E. 2011, ApJ, 726, 10

\bibitem[1999]{l1999} Livio M., Ogilvie G.~I., Pringle J.~E., 1999, ApJ,
512, 100

\bibitem[2003]{l2003} Livio, M., Pringle, J. E., \& King, A. R. 2003, ApJ, 593, 184

\bibitem[1994]{l1994} Lubow S. H., Papaloizou J. C. B., Pringle J. E., 1994, MNRAS, 267, 235

\bibitem[2012]{m2012} McKinney J. C., Tchekhovskoy A., \& Blandford R. D., 2012, MNRAS, 423, 3083

\bibitem[2000]{m2000} Meyer, F., Liu, B. F., \& Meyer-Hofmeister, E. 2000, A\&A, 361, 175


\bibitem[1995]{m1995} Murray, N., Chiang, J., Grossman, S. A., \& Voit, G. M. 1995, ApJ, 451, 498

\bibitem[2003]{n2003} Narayan R., Igumenshchev I. V., \& Abramowicz M. A., 2003, PASJ, 55, L69

\bibitem[1994]{n1994} Narayan R, \& Yi I. 1994, ApJ, 428, L13

\bibitem[2014] {Nixon2014} Nixon, C., \& Salvesen, G. 2014, MNRAS, 437, 3994

\bibitem[2011] {Orosz2011} Orosz, J. A., McClintock, J. E., Aufdenberg, J. P., et~al. 2011, ApJ, 742, 84

\bibitem[2012]{p2012} Ponti, G., Fender, R. P., Begelman, M. C., et al. 2012, MNRAS, 422, L11

\bibitem[2006]{r2006} Remillard, R. A., \& McClintock, J. E. 2006, ARA\&A, 44, 49

\bibitem[2003] {Ritter2003} Ritter, H., \& Kolb, U. 2003, A\&A, 404, 301

\bibitem[2008] {r2008} Rothstein, D. M., \& Lovelace, R. V. E. 2008, ApJ, 677, 1221

\bibitem[1973]{s1973} Shakura, N. I., \& Sunyaev, R. A. 1973, A\&A, 24, 337

\bibitem[2005]{s2005} Spruit, H. C., \& Uzdensky, D. A. 2005, ApJ, 629, 960


\bibitem[1989]{v1989} van Ballegooijen, A. A. 1989, in Astrophysics and Space Science Library, Vol. 156, Accretion Disks and Magnetic Fields in Astrophysics, ed. G., Belvedere (Dordrecht: Kluwer), 99

\bibitem[2007] {Yu2007} Yu, W., \& Dolence, J. 2007, ApJ, 667, 1043

\bibitem[2004] {Yu2004} Yu, W., van der Klis, M., \& Fender, R. 2004, ApJL, 611, L121

\bibitem[2009]{yu2009} Yu, W., \& Yan, Z. 2009, ApJ, 701, 1940

\bibitem[2009]{y2009} Yuan, F., Lin, J., Wu, K., \& Ho, L. C. 2009, MNRAS, 395, 2183

\bibitem[2012]{y2012} Yuan, F., Bu, D., Wu, M. 2012, ApJ, 761, 130

\bibitem[2014]{y2014} Yuan F., \& Narayan R., 2014, ARA\&A, 52, 529

\bibitem[2007] {Yuan2007} Yuan, F., Zdziarski, A. A., Xue, Y., \& Wu, X.-B. 2007, ApJ, 659, 541

\bibitem[2004] {Zdziarski2004} Zdziarski, A. A., Gierli{\'n}ski, M., Miko{\l}ajewska, J., et al. 2004, MNRAS, 351, 791

\bibitem[2013]{z2013} Zhang S., 2013, Frontier Phys., 8, 630

\end{thebibliography}
\end{document}